\documentstyle[prl,aps,multicol]{revtex}

\draft

\author{Tohya Hiroshima\thanks{Email: tohya@frl.cl.nec.co.jp}}
\title{Majorization Criterion for Distillability of a Bipartite Quantum State}
\address{Fundamental Research Laboratories, NEC Corporation,\\
34 Miyukigaoka, Tsukuba, 305-8501, Japan}

\begin{document}

\pagestyle{plain}
\pagenumbering{arabic}

\maketitle

\begin{abstract}
Bipartite quantum states are classified into three categories: separable states, bound entangled states, and free entangled states.
It is of great importance to characterize these families of states for the development of quantum information science.
In this paper, I show that the separable states and the bound entangled states have a common spectral property.
More precisely, I prove that for undistillable --- separable and bound entangled --- states, the eigenvalue vector of the global system is majorized by that of the local system.
This result constitutes a new sufficient condition for distillability of bipartite quantum states.
This is achieved by proving that if a bipartite quantum state satisfies the reduction criterion for distillability, then it satisfies the majorization criterion for separability.
\end{abstract}


\begin{multicols}{2}[]

\narrowtext

The recent development of quantum information science \cite{NC,BEZ,ABHHHRWWZ} has unveiled the rich structure of quantum states, in particular, the nature of quantum entanglement.
It is well acknowledged that the quantum entanglement is a physical resource in various types of quantum information processing such as quantum cryptography \cite{Ekert}, quantum dense coding \cite{BW}, quantum teleportation \cite{BBCJPW}, and quantum computation \cite{Shor}.
From the viewpoint of entanglement as a resource, bipartite quantum states are classified into three categories: separable states that are not entangled, bound entangled states, and free entangled states \cite{HHH98}.
This categorization is well understood through local quantum operations and classical communication (LOCC).
One of the most important LOCC protocol is the entanglement distillation or purification that allows us to extract pure maximally entangled states from several copies of a given free entangled state \cite{BBPSSW,BDSW,Gisin,DEMPS}.
However, the distillation protocol does not work for bound entangled states.
In spite of the practical importance of the distinction between free entangled states and bound entangled states, this distillability problem still remains open \cite{HH01,Bruss,BCHHKLS,Terhal}.

It is known that all states which violate the so-called reduction criterion \cite{HH99,CAG} are distillable.
The reduction criterion asserts that if a bipartite quantum state $\rho _{AB}$ on a composite Hilbert space ${\cal H}_{A}\otimes {\cal H}_{B}$ is undistillable, then the following operator inequalities are satisfied:
\begin{equation} \label{eq:reductionA}
\rho _{A}\otimes I_{B}\geq \rho _{AB}, 
\end{equation}
and
\begin{equation} \label{eq:reductionB}
I_{A}\otimes \rho _{B}\geq \rho _{AB},
\end{equation}
where $\rho _{A(B)}={\rm Tr}_{B(A)}\rho _{AB}$ is the reduction of $\rho _{AB}$, and $I_{A(B)}$ is the identity operator on ${\cal H}_{A(B)}$.
The reduction criterion is also a necessary and sufficient condition for separability in low dimensional composite states with $\dim {\cal H}_{A}=2$ and $\dim {\cal H}_{B}=2$ or $3$.
Recently, Nielsen and Kempe \cite{NK} proposed a new criterion for separability --- the majorization criterion which asserts that if
$\rho _{AB}$ is separable, then
\begin{equation} \label{eq:majorizationA}
\lambda (\rho _{AB})\prec \lambda (\rho _{A}), 
\end{equation}
and
\begin{equation} \label{eq:majorizationB}
\lambda (\rho _{AB})\prec \lambda (\rho _{B}), 
\end{equation}
where $\lambda (\rho _{AB})$ is a vector of eigenvalues of $\rho _{AB}$;
$\lambda (\rho _{A}) $ and $\lambda (\rho _{B})$ are defined similarly.
The relation $x\prec y$
between $n$-dimension vectors $x$ and $y$, which reads ``$x$ is majorized by $%
y $'', means that
\begin{equation} \label{eq:majdef1}
\sum_{i=1}^{k}x_{i}^{\downarrow }\leq \sum_{i=1}^{k}y_{i}^{\downarrow
}\qquad (1\leq k\leq n-1), 
\end{equation}
and
\begin{equation} \label{eq:majdef2}
\sum_{i=1}^{n}x_{i}^{\downarrow }=\sum_{i=1}^{n}y_{i}^{\downarrow }, 
\end{equation}
where $x_{i}^{\downarrow }$ $(1\leq i\leq n)$ are components of vector $%
x $ rearranged in decreasing order ($x_{1}^{\downarrow }\geq
x_{2}^{\downarrow }\geq \cdots \geq x_{n}^{\downarrow }$); $y_{i}^{\downarrow
}$ $(1\leq i\leq n)$ are defined similarly. If the dimensions of $x$ and $y$ are different,
the smaller vector is enlarged by appending extra zeros to
equalize their dimensions \cite{Nielsen}.
In Eqs.~(\ref{eq:majorizationA}) and (\ref{eq:majorizationB}), $\lambda (\rho _{A})$ and $\lambda (\rho _{B})$ are considered enlarged vectors with dimensions the same as that of $\lambda (\rho _{AB})$.
The majorization criterion has an intuitive physical interpretation; the separable states are more disordered globally than locally, as stated in the title of Ref.~\cite{NK}.

Now a question arises: in which ways are these two (reduction and
majorization) criteria related? It has been conjectured that the
majorization criterion is implied by the reduction crierion, but this has not been proven \cite{Bruss,Terhal,VW}.
In this paper, I prove that this conjecture is true. Furthermore, from this result I propose a new criterion for distillability.
As for the first result, I report the following theorem.

{\em Theorem 1:} If $\rho _{AB}$ is a density matrix
such that $\rho _{A}\otimes I_{B}\geq \rho _{AB}$,
then $\lambda (\rho _{AB})\prec \lambda (\rho _{A})$.

Before proving Theorem 1, I will present two lemmas and the generalization of the majorization concept.

Let $A$ and $B$ be hermitian operators acting on a finite dimensional
Hilbert space. The following holds.

{\em Lemma 1:} If $0\leq A\leq B$ and $B>0$, then there exists an operator $C$ such that $%
A^{1/2}=B^{1/2}C$ and $\left\| C\right\| _{\infty }\leq 1$. Here, $\left\|
C\right\| _{\infty }$ is the operator norm of $C$ and is defined as $\left\|
C\right\| _{\infty }=\sup_{\left\| x\right\| =1}\left\| Cx\right\| $.

This lemma is a weak version of Douglas' theorem \cite{Douglas}: (i)
the inequality $AA^{\dagger }\leq BB^{\dagger }$ holds if and only if there
exists an operator $C$ such that $A=BC$ and (ii) if (i) is valid, then
there exists a unique $C$ such that $\left\| C\right\| _{\infty }\leq 1$.
Although Douglas' original proof is mathematically
sophisticated, I can show a very simple proof of Lemma 1.

{\em Proof of Lemma 1:} Since $B>0$, $B^{-1/2}$ is well
defined and the inequalities $0\leq A\leq B$ are equivalent to $I\geq B^{-1/2}AB^{-1/2}=CC^{\dagger }\geq 0$ with $C=B^{-1/2}A^{1/2}$.
Hence, $\left\| C\right\|_{\infty }\leq 1$ (e.g., Lemma V.1.7 in Ref.~\cite{Bhatia}).
Therefore, the proof of Lemma 1 is completed. $\Box$

{\em Lemma 2:} A bipartite density matrix $\rho _{AB}$ on ${\cal H}_{A}\otimes {\cal H}_{B}$ is written as $\rho _{AB}=\rho _{AB}^{\prime }\oplus 0$.
Here, the zero operator $0$ acts on ${\rm Ker}(\rho _{A})\otimes {\cal H}_{B}$ and $\rho _{AB}^{\prime }$ acts on ${\rm Ker}(\rho _{A})^{\perp }\otimes {\cal H}_{B}$, i.e., the orthogonal compliment of ${\rm Ker}(\rho _{A})\otimes {\cal H}_{B}$, where ${\rm Ker}(\rho _{A})$ is the kernel of $\rho _{A}$ defined as ${\rm Ker}(\rho _{A})=\left\{ \left| \psi \right\rangle \in {\cal H}_{A} ;\rho _{A}\left| \psi \right\rangle =0\right\} $.

Note that $\rho _{AB}^{\prime }$ is invertible density matrix in the
restricted subspace ${\rm Ker}(\rho _{A})^{\perp }\otimes {\cal H}_{B}$.
The following proof is due to Audenaert \cite{A}.

{\em Proof of Lemma 2:} Let $\left| x\right\rangle $ ($\left| y\right\rangle $) be a state vector in ${\rm Ker}(\rho _{A})$ (${\cal H}_{B}$).
Since $\left| y\right\rangle
\left\langle y\right| \leq I_{B}$ and $\left| x\right\rangle \left\langle
x\right| ,\rho _{AB}\geq 0$, we have
\begin{eqnarray}
0 &\leq &\left\langle x,y\right| \rho _{AB}\left| x,y\right\rangle
\leq{\rm Tr}_{AB}\left[ \left( \left| x\right\rangle \left\langle x\right| \otimes
I_{B}\right) \rho _{AB}\right]  \nonumber \\
&=&\left\langle x\right| \rho _{A}\left| x\right\rangle =0
\end{eqnarray}
so that $\rho _{AB}\left| x,y\right\rangle =0$. Therefore, $\rho _{AB}\left|
\psi \right\rangle =0$ for every state vector $\left| \psi \right\rangle \in
{\rm Ker}(\rho _{A})\otimes {\cal H}_{B}$ because $\left| \psi \right\rangle $ is written
as a superposition of $\left| x,y\right\rangle $ with $\left| x\right\rangle
\in {\rm Ker}(\rho _{A})$ and $\left| y\right\rangle \in {\cal H}_{B}$. That is, $\rho
_{AB}=0$ on ${\rm Ker}(\rho _{A})\otimes {\cal H}_{B}$. This completes the proof of Lemma 2. $\Box$

Now I will describe the concept of ``weak'' majorization.
If the last equality [Eq.~(\ref{eq:majdef2})] is also an inequality
\begin{equation}
\sum_{i=1}^{n}x_{i}^{\downarrow }\leq \sum_{i=1}^{n}y_{i}^{\downarrow }, 
\end{equation}
$x$ is said to be weakly submajorized by $y$. The symbol ``$\prec $'' is now
written as ``$\prec _{w}$'': $x\prec _{w}y$ . The necessary and sufficient
condition for the relation $x\prec _{w}y$ is that there exists an $n$ by $n$
doubly substochastic matrix $S$ such that $x=Sy$.
The proof of this proposition can be found in standard textbooks on matrix theory \cite{Bhatia,MO,HJ2}.
Here, an $n$ by $n$ real matrix $S$ is said to be doubly substochastic if
(i) the entries in $S$ are nonnegative; $S_{i,j}\geq 0$;
(ii) all row sums of $S$ are at most one; $\sum_{j=1}^{n}S_{i,j}\leq 1$ ($%
1\leq i\leq n$); 
and
(iii) all column sums of $S$ are at most one; $\sum_{i=1}^{n}S_{i,j}\leq 1$ (%
$1\leq j\leq n$).

If the inequalities in (ii) and (iii) are replaced by corresponding
equalities, $S$ is said to be doubly stochastic. The existence of a
doubly stochastic matrix such that $x=Sy$ is equivalent to the usual majorization relation $%
x\prec y$.

{\em Proof of Theorem 1:} By virtue of Lemma 2,
we can assume that $\rho _{A}$ is invertible without loss of generality.
Therefore, by Lemma 1, Eq.~(\ref{eq:reductionA}) implies the existence of
an operator $R$ such that
\begin{equation} \label{eq:douglas}
\rho _{AB}^{1/2}=\left( \rho _{A}^{1/2}\otimes
I_{B}\right) R,
\end{equation}
with $\left\| R\right\| _{\infty }\leq 1$.
It is also assumed that $\rho _{A}$ is diagonal;
\begin{equation}
\rho _{A}={\rm diag} \left( \lambda _{1}(\rho _{A}),\lambda _{2}(\rho _{A}),\cdots
,\lambda _{d_{A}}(\rho _{A})\right) \equiv {\rm diag}\lambda (\rho _{A}).
\end{equation}
Here and hereafter, both double and
single indexing are used interchangeably to indicate entries in matrices and
in vectors of the composite system $AB$. The double indices are enclosed in square brackets. As an example, suppose $M$ is a matrix acting on the composite
space ${\cal H}_{A}\otimes {\cal H}_{B}$. The matrix elements are usually written on some
product basis such as $M_{12,34}=\left\langle 1_{A}\right| \otimes
\left\langle 2_{B}\right| M\left| 3_{A}\right\rangle \otimes \left|
4_{B}\right\rangle $, where $\left| i _{A(B)}\right\rangle$ ($1\leq i\leq
d_{A(B)}$) forms an orthogonal basis in ${\cal H}_{A(B)}$.
Instead of using this conventional notation,
$M_{12,34}$ is written as $M_{[1,2],[3,4]}$ or $M_{[1,2],(3-1)d_{B}+4}$ in the following.
This notation
makes the following calculations unequivocal and easier to follow. Since $\rho
_{AB}^{1/2}$ is hermitian, it is diagonalized by a suitable unitary operator 
$V$:
\begin{eqnarray} \label{eq:diagAB}
V^{\dagger }\rho _{AB}^{1/2}V &=&{\rm diag}\left( \sqrt{\lambda _{1}(\rho _{AB})},%
\cdots ,\sqrt{\lambda _{d_{A}d_{B}}(\rho _{AB})}\right)  \nonumber \\
&\equiv &{\rm diag}\sqrt{\lambda (\rho _{AB})}.
\end{eqnarray}
Here, $\lambda _{i}(\rho _{AB})$ ($1\leq i\leq d_{A}d_{B}$) are eigenvalues of $\rho _{AB}$ and are ordered decreasingly so that $\lambda (\rho _{AB})=\lambda
^{\downarrow }(\rho _{AB})$ without loss of generality. Note that ${\rm Tr}_{B}\left(
V^{\dagger }\rho _{AB}V\right) \neq \rho _{A}$ in general.
From Eqs.~(\ref{eq:douglas}) and (\ref{eq:diagAB}), we have 
\begin{equation} \label{eq:eq0}
{\rm diag}\sqrt{\lambda (\rho _{AB})}=V^{\dagger }\left( \rho _{A}^{1/2}\otimes
I_{B}\right) C,
\end{equation}
where $C=RV$, and it is also a contraction; $\left\| C\right\| _{\infty }\leq
\left\| R\right\| _{\infty }\left\| V\right\| _{\infty }\leq 1$, i.e., the
maximum eigenvalue of $C^{\dagger }C$ is at most one. Since the diagonal
elements of a hermitian matrix do not exceed its maximum eigenvalue \cite{Bhatia,MO,HJ1},
$\left( C^{\dagger }C\right) _{[i,j],[i,j]}\leq 1$, i.e.,
\begin{equation} \label{eq:constraint1}
\sum_{k=1}^{d_{A}}\sum_{l=1}^{d_{B}}\left| C_{[k,l],[i,j]}\right| ^{2}\leq
1 
\end{equation}
for $1\leq i\leq d_{A}$ and $1\leq j\leq d_{B}$.
Now, from Eq.~(\ref{eq:eq0}) we have
\begin{equation} \label{eq:eq1}
\rho _{AB}=\left( \rho _{A}^{1/2}\otimes I_{B}\right) CC^{\dagger }\left(
\rho _{A}^{1/2}\otimes I_{B}\right),
\end{equation}
and
\begin{equation} \label{eq:eq2}
{\rm diag} \lambda (\rho _{AB})=C^{\dagger }\left( \rho _{A}\otimes I_{B}\right) C.
\end{equation}
The diagonal elements of Eq.~(\ref{eq:eq1}) yield
\begin{eqnarray}
\lambda _{i}(\rho _{A}) &=&\left( \rho _{A}\right) _{i,i}=\sum_{j=1}^{d_{B}}\left(
\rho _{AB}\right) _{[i,j],[i,j]} \nonumber \\
&=&\lambda
_{i}(\rho _{A})\sum_{j=1}^{d_{B}}\sum_{k=1}^{d_{A}}\sum_{l=1}^{d_{B}}\left|
C_{[i,j],[k,l]}\right| ^{2}.
\end{eqnarray}
Since $\rho _{A}$ is invertible, all eigenvalues of $\rho _{A}$ are strictly
positive: $\lambda _{i}(\rho _{A})>0$ ($1\leq i\leq d_{A}$).
Hence,
\begin{equation} \label{eq:constraint2}
\sum_{j=1}^{d_{B}}\sum_{k=1}^{d_{A}}\sum_{l=1}^{d_{B}}\left|
C_{[i,j],[k,l]}\right| ^{2}=1.
\end{equation}
Equations (\ref{eq:constraint1}) and (\ref{eq:constraint2}) constitute the constraints on the entries of $C$.
To derive a linear equation between $\lambda (\rho _{AB})$ and $\lambda (\rho _{A})$, we use Eq.~(\ref{eq:eq2}).
The diagonal elements of this equation yield
\begin{equation}
\lambda _{[i,j]}(\rho _{AB}) =\sum_{k=1}^{d_{A}}\sum_{l=1}^{d_{B}}\lambda
_{k}(\rho _{A})\left| C_{[k,l],[i,j]}\right| ^{2}.
\end{equation}
Namely,
\begin{eqnarray}
&&\left( \lambda _{1}(\rho _{AB}),\lambda _{2}(\rho _{AB}),\cdots ,\lambda _{d_{A}}(\rho _{AB})\right)
^{t} \nonumber \\
&=&S\left( \lambda _{1}(\rho _{A}),\lambda _{2}(\rho _{A}),\cdots ,\lambda _{d_{A}}(\rho _{A})\right)
^{t}.
\end{eqnarray}
Here, the $d_{A}$ by $d_{A}$ matrix $S$ is defined as
\begin{equation} \label{eq:matrix1}
S_{i,j}=\sum_{k=1}^{d_{B}}\left| C_{[j,k],i}\right| ^{2} \geq 0 \qquad (1\leq i,j\leq d_{A}).
\end{equation}
The row sum of the $i$-th row of $S$ is calculated as
\begin{equation} \label{eq:matrix2}
\sum_{j=1}^{d_{A}}S_{i,j}=\sum_{j=1}^{d_{A}}\sum_{k=1}^{d_{B}}\left|
C_{[j,k],i}\right| ^{2}\leq 1.
\end{equation}
The last inequality is due to Eq.~(\ref{eq:constraint1}).
The column sum of the $j$-th column of $S$ is calculated as
\begin{equation} \label{eq:matrix3}
\sum_{i=1}^{d_{A}}S_{i,j}=\sum_{i=1}^{d_{A}}\sum_{k=1}^{d_{B}}\left|
C_{[j,k],i}\right| ^{2}\leq \sum_{i=1}^{d_{A}d_{B}}\sum_{k=1}^{d_{B}}\left|
C_{[j,k],i}\right| ^{2}=1. 
\end{equation}
The last equality follows from Eq.~(\ref{eq:constraint2}).
From Eqs.~(\ref{eq:matrix1}), (\ref{eq:matrix2}), and (\ref{eq:matrix3}), $S$ is doubly substochastic.
Hence,
\begin{eqnarray}
&&\left( \lambda _{1}(\rho _{AB}),\lambda _{2}(\rho _{AB}),\cdots ,\lambda _{d_{A}}(\rho _{AB})\right) 
\nonumber \\
&\prec _{w}& \left( \lambda _{1}(\rho _{A}),\lambda _{2}(\rho _{A}),\cdots ,\lambda
_{d_{A}}(\rho _{A})\right) .
\end{eqnarray}
Since $\lambda _{i}(\rho _{AB})$ ($1\leq i\leq d_{A}$) are the first $d_{A}$ largest
eigenvalues of $\rho _{AB}$, we can conclude that
\begin{equation}
\sum_{i=1}^{k}\lambda _{i}(\rho _{AB})\leq \sum_{i=1}^{k}\lambda _{i}(\rho _{A}) \qquad (1\leq k\leq d_{A}d_{B}) 
\end{equation}
with the inequality holding equality for $k=d_{A}d_{B}$ due to the obvious
fact that ${\rm Tr}_{AB}\rho _{AB}={\rm Tr}_{A}\rho _{A}=1$.
Since this final conclusion is equivalent to the majorization relation, $\lambda (\rho _{AB})\prec \lambda (\rho _{A})$, the proof of
Theorem 1 is completed. $\Box$

The converse of Theorem 1 is not generally true.
There is a conterexample of the maximally entangled mixed state \cite{IH} with rank two (Example 1 in Ref.~\cite{NK}).
However, there exist some families of states for which the majorization criterion detects their entanglement perfectly.
The isotropic states in arbitrary dimensions belong to such examples \cite{NK}.

Theorem 1 is also connected with the distillability problem.
By Theorem 1 together with the fact that all states which cannot be distilled
satisfy the reduction criterion \cite{HH99}, we immediately arrive at the following theorem.

{\em Theorem 2:} If $\rho _{AB}$ is not distillable, then $\lambda (\rho _{AB})\prec \lambda (\rho _{A})$ and $\lambda (\rho _{AB})\prec \lambda (\rho _{B})$.

Equivalently, all states which violate the majorization criterion are distillable.
This constitutes a new sufficient condition for distillability of bipartite states.
As an example, a family of maximally correlated states \cite{Rains} of the form
\begin{equation}
\rho _{AB}=\sum_{i,j}\alpha _{ij}\left| i_{A}\right\rangle \left\langle j_{A}\right| \otimes \left|i_{B}\right\rangle \left\langle j_{B}\right|
\end{equation}
violates the majorization criterion so that it is distillable except when all $\alpha _{ij}=0$ for $i\neq j$.
Since $\rho _{A}=\sum_{i}\alpha _{ii}\left| i_{A}\right\rangle \left\langle i_{A}\right| $, the eigenvalues of $\rho _{A}$ are exactly the diagonal elements
of $\rho _{AB}$. Therefore, $\lambda (\rho _{A})\prec \lambda (\rho _{AB})$ because the vector of diagonal elements in a hermitian matrix is majorized by that of its eigenvalues \cite{Bhatia,MO,HJ1}.
Furthermore, it is evident from Theorem 2 that a bound entangled state shares a common spectral property with a separable state.
Namely, for an undistillable (separable and bound entangled) state $\rho _{AB}$ the global spectra $\lambda(\rho _{AB})$ is majorized by the local spectra $\lambda(\rho _{A})$ and $\lambda(\rho _{B})$.

In conclusion, the problem of relating the reduction criterion for distillability with the majorization criterion for separability has been finally solved.
That is, if a bipartite quantum state satisfies the reduction criterion, then it satisfies the majorization one as well.
From this result, I have found that for a bound entangled state as well as a separable state the eigenvalue vector of the global system is majorized by that of the local (reduced) system.
Furthermore, a new sufficient condition for distillbilty of a bipartite state has been proposed.
I hope that these new results trigger the discovery of new distillation protocols and also stimulate the progress on the theory of quantum entanglement.

I would like to thank Koenraad M. R. Audenaert for enlightening me on the
fact stated in Lemma 2 as well as for helpful comments.
This work was supported by CREST of the Japan Science and Technology Corporation (JST).

\end{multicols}

\end{document}